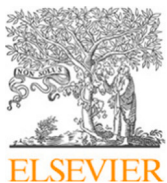
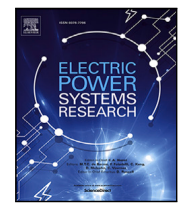

# Multi-point method using effective demodulation and decomposition techniques allowing identification of disturbing loads in power grids

Piotr Kuwałek *, Grzegorz Wiczyński

Institute of Electrical Engineering and Electronics of the Poznan University of Technology, Poznan, Poland

A B S T R A C T

The paper presents an innovative approach to identifying voltage fluctuation sources in power networks, also considering the localization understood as the indication of supply points of disturbing loads. The presented approach considers disturbance sources that change their operating state with a frequency higher than the power frequency. The implementation of the proposed solution is also proposed in such a way that its implementation in the smart meter infrastructure allows automatic localization of disturbance sources without additional expert knowledge. In the proposed approach, the modulation signal is estimated using a carrier signal estimator, which allows the estimation of a modulation signal with a frequency higher than the power frequency. The estimated modulating signal is decomposed into component signals associated with individual disturbing loads by decomposition by approximation using pulse waves. The decomposition process allows for the estimation of selected parameters associated with disturbing loads, on the basis of which the assessment of propagation of voltage fluctuations associated with the impact of individual disturbance sources is performed, which allows for the indication of their supply point. The proposed approach was verified in numerical simulation studies using MATLAB/SIMULINK and in experimental studies carried out in a real low-voltage power grid.

**Nomenclature**

| | |
|---|---|
| $A_i$ | Estimated amplitudes of $i$–th modulating signal component in period of discrimination (contribution of the $i$–th disturbing load to the resultant voltage fluctuations). |
| $A_{i\max}$ | Maximum value of estimated amplitudes of $i$–th modulating signal component in period of discrimination. |
| $A_i^{P_j}$ | Estimated amplitudes of $i$–th modulating signal component in period of discrimination determined for the $j$th supply point. |
| AM | Amplitude modulation. |
| AP | Active probe. |
| DAPW | Decomposition by approximation with pulse waves. |
| $e_1$ | Accuracy metric to determine the effectiveness of the correct indication of the supply point of the source of disturbance. |
| $e_2$ | Accuracy metric to determine the effectiveness of the correct estimation of the frequency of changes in the operating state of the source of disturbance. |
| EEWT | Enhanced empirical wavelet transform. |
| $f_c$ | Power frequency (50 Hz or 60 Hz) - nominal voltage frequency (carrier frequency). |
| $f_i$ | Estimated fundamental frequency of $i$–th modulating signal component in a period of discrimination (frequency associated with the rate of changes in the operating state of the $i$–th disturbing load). |

☆ This work was funded by National Science Centre, Poland – 2021/41/N/ST7/00397. For the purpose of Open Access, the author has applied a CC-BY public copyright licence to any Author Accepted Manuscript (AAM) version arising from this submission.
* Corresponding author.
 *E-mail address:* piotr.kuwalek@put.poznan.pl (P. Kuwałek).

https://doi.org/10.1016/j.epsr.2024.110335
Received 7 February 2024; Received in revised form 6 March 2024; Accepted 16 March 2024
Available online 21 March 2024

| | |
|---|---|
| $f_i^{P_j}$ | Estimated fundamental frequency of $i$–th modulating signal component in period of discrimination determined for the $j$th supply point. |
| $l$ | Distance from power station. |
| $l_i$ | Length of $i$–th section of power line. |
| $l_{P_i}$ | Distance of $i$–th supply point $P_i$ from power station. |
| $L_i$ | Inductance of $i$th section of the power line. |
| LV | Low-voltage. |
| M1 | Method based on the evaluation of the propagation of voltage fluctuations using nonparametric statistical evaluation of the voltage fluctuations indices without additional expert knowledge. |
| M2 | Method based on the evaluation of flicker propagation. |
| M3 | Method based on decomposition using enhanced empirical wavelet transform. |
| MOSFET | Metal–oxide–semiconductor field-effect transistor. |
| MV | Medium-voltage. |
| $N$ | Number of components of voltage modulating signal obtained using DAPW (number of significant disturbing loads). |
| $N_{f_T}$ | Number of correctly identified frequencies of changes in the operating state of individual disturbing loads. |
| $N_T$ | Number of correctly indicated supply points (localized) disturbing loads. |
| $P_i$ | $i$–th supply point of loads in the power grid. |
| $P_{lt}$ | Long-term flicker indicator. |
| $P_{st}$ | Short-term flicker indicator. |
| $P_{N_i}$ | Nominal value of active power of individual sources of disturbance $i$–th supply point of loads in the power grid. |
| PA | Proposed approach. |
| PQ | Power quality. |
| $R$ | Resistance. |
| $R_i$ | Resistance of $i$–th section of power line. |
| SSR | Solid state relay. |
| THD | Total harmonic distortion. |
| $u(t)$ | Instantaneous voltage values. |
| $u_{mod_i}$ | Estimated $i$–th component signal associated with the influence of the $i$–th disturbing load. |
| $u_{P_i}(t)$ | Instantaneous voltage values at point $P_i$. |
| $U$ | Root-mean-square (rms) value of voltage. |
| $U_m$ | Amplitude of voltage. |
| $X$ | Reactance. |
| $Z$ | Impedance. |

## 1. Introduction

Power quality (PQ) is an important topic related to modern power systems. In many countries, new legal documents (e.g. [1]) appear forcing power suppliers to ensure that the electrical energy they supply is of appropriate quality. PQ is mainly a set of parameters that determine the quality of the supply voltage [2], e.g.: rms value of voltage $U$, total harmonic ratio THD, contribution of individual harmonics, short-term $P_{st}$ and long-term $P_{lt}$ flicker indicator. Parameters that determine PQ allow for the evaluation of the severity of PQ disturbances in power networks. One of the most common disturbances in power grids is voltage fluctuations [3], which can influence the operation of electrical loads such as motors [4–6] or light sources [7,8]. In the case of light sources, flicker can occur [8–10], which in bad conditions can induce epileptic or depressive states. The scale of the problem of voltage fluctuations increases with the development trend of modern power systems [11], which are focused on increasing the share of renewable energy sources in the manufacturing sector [12–14] and increasing the number of power electronic loads [15,16], e.g., working with renewable energy sources, or used in energy-saving solutions, or used in control systems. Most often, the problem of voltage fluctuations is solved when a consumer reports a complaint, and then actions are taken to identify disturbing loads in the power grid, which in turn are focused on minimizing the disruptions caused by these disturbing loads. Such actions include, for example, the reconfiguration of the distribution system [17], adaptive voltage control [18], selective interharmonic compensation by adding a selective interharmonic filter to the classical Statcom control [19], installation of a shunt active filter [20] or other specialized compensators [21], modifications of system solutions in the architecture used [22] or modifications of converter systems [23].

Methods for identification of sources of voltage fluctuations can be divided into single-point methods (e.g., the correlation of changes in the flicker severity $P_{st}$ and/or changes in the active and reactive power [24], the analysis of voltage and current measurements and the examination of correlation coefficients [25], the identification of interharmonic power direction [26], the examination of the voltage fluctuation power [27], flicker responsibility division method based on the flicker power [28], the assessment of correlation of change of power and tension in a power supply network [29]) and multi-point methods (e.g., the statistical analysis of voltage fluctuation indices [30], the statistical analysis of voltage fluctuation indices, $P_st$ indicator and current changes [31], using multi level perceptron neural network for bus flickr indices [32], the assessment of the gradient of voltage amplitudes and constructing a Jacobian matrix [33], selected normative conventional methods [34], analysis of the frequency–amplitude binary characterization for individual network points [35], the evaluation of the flicker propagation using a short-circuit-based method for determining the flicker transfer coefficient [36], the method using the wavelet transform, the approximate model of the line and flicker power theory [37]). Single-point methods have limited diagnostic capabilities because they only allow for identification of the side responsible for voltage fluctuations from the perspective of the measurement point. If a specific measurement point is not the supply point of the disturbing load, identification of the disturbing load should be performed iteratively at subsequent points in the power grid. However, in the case of many sources of voltage fluctuations, including power electronic loads that change their operating state at a frequency higher than the power frequency $f_c$, single-point methods often cause incorrect conclusions about the side that is the source of disturbance [26,38]. Multi-point methods have greater diagnostic capabilities, but require more measurements and require greater computational complexity. However, it is worth noting the numerous limitations of the multi-point methods currently available in the literature [31,33,38]. First, most of the multi-point methods available in the literature do not allow the identification of disturbing loads that change their operating state at a frequency higher than the power frequency $f_c$ (e.g. due to the limitations of voltage fluctuation indices determined from changes in the rms value of voltage in the method [30] or [31], due to the method of determination of the flicker in the method [32] or [33], due to the averaging nature of the rms value and flicker evaluation methods [34]). In the case of multi-point methods based on machine learning methods [39–41] (including deep learning methods [42,43]), effective operation requires appropriately prepared training data, which are difficult to obtain due to the constantly changing nature of the loads in the network. Even if identification is limited to loads that change their operating state at a frequency lower than the power frequency $f_c$, the multi-point methods currently available in the literature provide limited identification possibilities (e.g., some methods only provide the ability to indicate the supply point of disturbing loads [30,31] or to estimate selected parameters associated with the impact of a certain group loads, e.g., frequency of changes in the status of individual sources of disturbance [44]). In recent years, the first proposal of a multi-point method based on an enhanced empirical wavelet transform [38], which is the first method to allow preliminary selective identification that result in the indication of the point of supply of the source of disturbance and the estimation of selected parameters associated with individual sources of voltage fluctuations, including disturbing loads changing their operating state with a frequency higher than the power frequency $f_c$. However, due to the limitations of the enhanced empirical wavelet transform method, it was shown in [45] that the approach based on this algorithm under bad conditions also causes an incorrect indication of the supply point of the disturbing load, or an incorrect assessment of selected features of disturbing loads. Therefore, this article presents a modification of the method described in [38], which does not have the limitations indicated in [45], and therefore solves the problem of effective identification of sources of voltage fluctuations in power grids. Furthermore, the proposed method implemented in the smart energy meter infrastructure creates the possibility of automatic identification, which does not require additional expert knowledge. The proposed approach allows both the indication of the power supply point of a disturbing load and the estimation of selected parameters associated with individual disturbing loads at individual points in the power grid. The selected parameters include the frequency of changes in the operating state of disturbance sources (a feature dependent only on the disturbing load) and the average amplitude of voltage fluctuations at individual points in the power grid caused by a specific source of disturbance (a feature dependent on the disturbing load and on the parameters of the supply circuit). The correctness of the proposed approach was verified for a large set considering many power grid topologies and the changing nature of disturbing loads (along with a different number of their simultaneous occurrences) obtained from numerical simulation studies and experimental studies.

To sum up, the novelty of the paper includes:

- presentation of a prepared new method for identification of voltage fluctuations sources in a power network with a branching radial topology;
- obtaining the ability to effectively identify sources of voltage fluctuations that change their operating state with a frequency higher than the power frequency $f_c$, for the prepared method, by using an innovative signal chain cascade: demodulation by [46] – decomposition by [47] – statistical assessment of propagation;
- obtaining the ability to effectively identify sources of voltage fluctuations that have a similar frequency of operating state changes for the prepared method by using the proprietary decomposition method [47];
- validation of the proposed approach on a specially prepared data set, which includes results for a modern power grid with simultaneous PQ disturbances (unlike other solutions available in the literature, which were often validated for idealized conditions that do not occur in modern power grids) obtained in numerical simulation studies and laboratory experimental studies, and which is made publicly available as a baseline [48];
- obtaining the possibility of automatic identification of sources of voltage fluctuations without additional expert knowledge, in the case of implementation of the proposed approach in the infrastructure of smart energy meters.

## 2. Proposed approach

The proposed algorithm for the identification of sources of voltage fluctuations is presented in Fig. 1. The idea of the proposed approach uses the assessment of the propagation of voltage fluctuations in the power grid [30,31]. An example of the propagation of voltage fluctuations is shown in Fig. 2. Assessment of the propagation of voltage fluctuations consists in assessment of the amplitude of voltage fluctuations for a specific source of voltage fluctuations. The amplitude of voltage fluctuations in the power grid increases to its maximum value at the supply point of the disturbing load. If there is no propagation of voltage fluctuations on the medium-voltage MV side, then on the low-voltage LV side, the principle can be that the closer to the MV/LV supply point, the lower the amplitude of voltage fluctuations. For supply points further away from the MV/LV power supply point and the supply point of the disturbance source, it can be seen that the amplitude of voltage fluctuations is constant or its slight attenuation is visible in the case of the capacitive nature of the line. In the case of the propagation of voltage fluctuations from the MV side, the assessment of the propagation of voltage fluctuations is analogous, except that it is assumed that the MV/LV power supply point is treated as an additional supply point for the disturbance source. Currently, in the literature, methods based on the assessment of the propagation of voltage fluctuations allow the effective identification of the dominant source of disturbance and are limited to the identification of sources of disturbance that change their operating state at a frequency lower than the power frequency $f_c$. Therefore, to increase the diagnostic capabilities of the proposed approach, the assessment of the propagation of voltage fluctuations in this approach is performed for selected parameters already associated with the impact of individual (specific) sources of disturbances in the power grid. The selected parameters are the frequency of changes in the operating state of individual sources of disturbance (the feature/parameter depends only on the source of disturbance) and the amplitude of voltage changes induced by individual sources of disturbance (feature/parameter depending on the source of disturbance and the properties of the supply circuit). The amplitude of voltage changes induced by individual sources of disturbance can also be treated as the contribution of a specific source of disturbance to the resulting voltage fluctuations. In order to estimate selected parameters associated with the impact of individual disturbing loads, in the first step the voltage signal from individual points of the network is demodulated using a demodulator with carrier signal estimation, which is described in detail in [46]. At this stage, a modulating signal is obtained that determines the resulting voltage fluctuations in the power grid. In the second step, the modulating signal is decomposed into component signals using decomposition by approximation with pulse waves (DAPW), which is described in detail in [47]. The proposed use of the developed decomposition method in the process of identifying disturbing loads allows an effective average assessment of the contribution of individual disturbing loads to the resulting voltage fluctuations. In the method described in [47], the actual shapes of voltage fluctuations are considered, which allow the decomposition method to work effectively without additional conditions that had to be applied in the proposal described in [38], where the enhanced empirical wavelet transform (EEWT) method is used for decomposition. The EEWT method, even with additional conditions imposed in bad conditions, causes incorrect identification of sources of voltage fluctuations (e.g. incorrect indication of the supply point of the source of disturbance, incorrect estimation of the frequency of changes in the operating state of the identified source of disturbance) [45].

Hence, the newly proposed approach based on demodulation-decomposition-propagation assessment, where DAPW is used for decomposition, solves the problem of effective identification of sources of voltage fluctuations in the power grid. The conceptual operation of the proposed approach is shown in Fig. 3. It should be noted that the proposed approach proposes the task of identification of sources of disturbance when the limit value is exceeded by the selected threshold parameter. Such activation of the identification procedure supports the implementation of the approach in the currently implemented smart meter infrastructure, or in the measurement and recording infrastructure based on Internet of Things technologies. For the purposes of the publication, it is proposed to adopt the short-term flicker indicator $P_{st}$ as a threshold parameter, because currently this indicator is used to assess voltage fluctuations in power grids. It is proposed to adopt a value equal to 1.0 as the limit value at any point in the power grid, because it is the limit value in LV networks specified in normative documents [2].

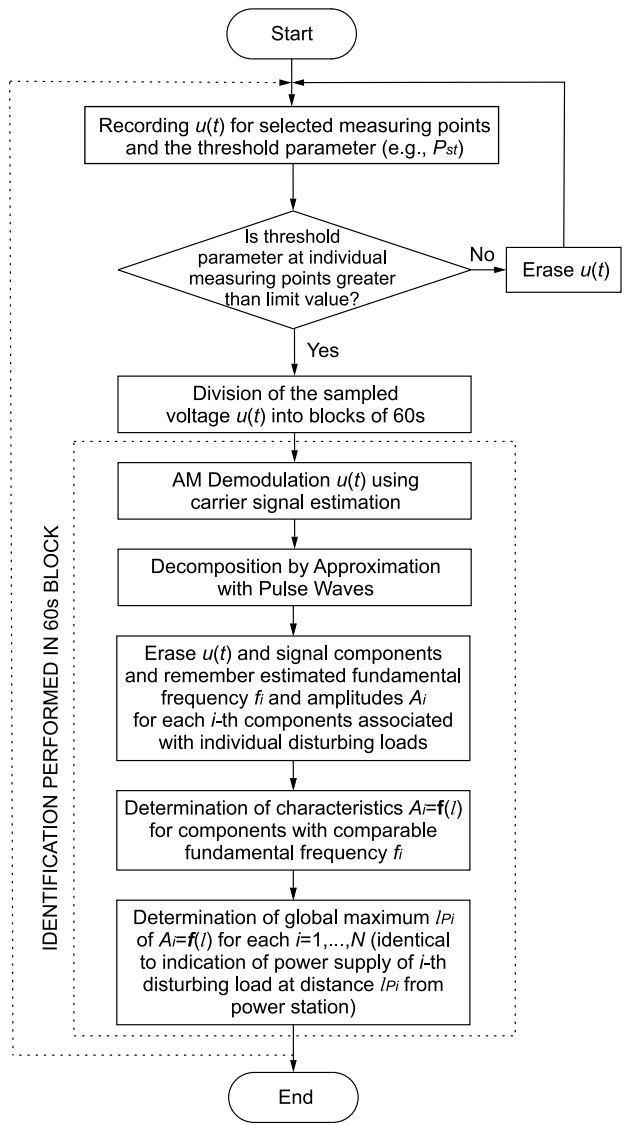

Fig. 1. The proposed algorithm for selective identification and localization of individual disturbing load.

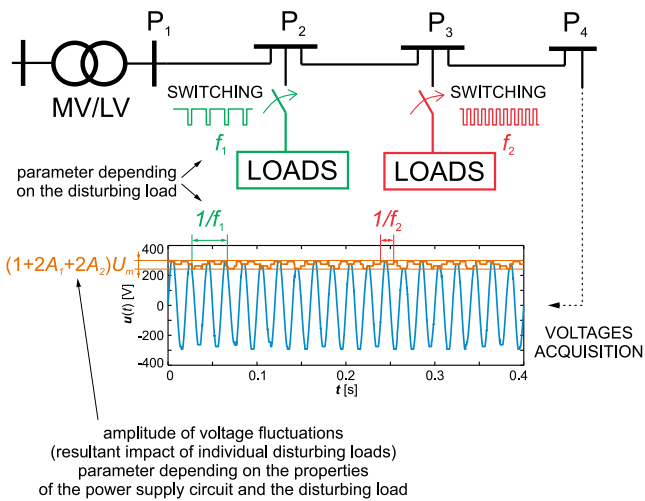

Fig. 2. An example of the propagation of voltage fluctuations.

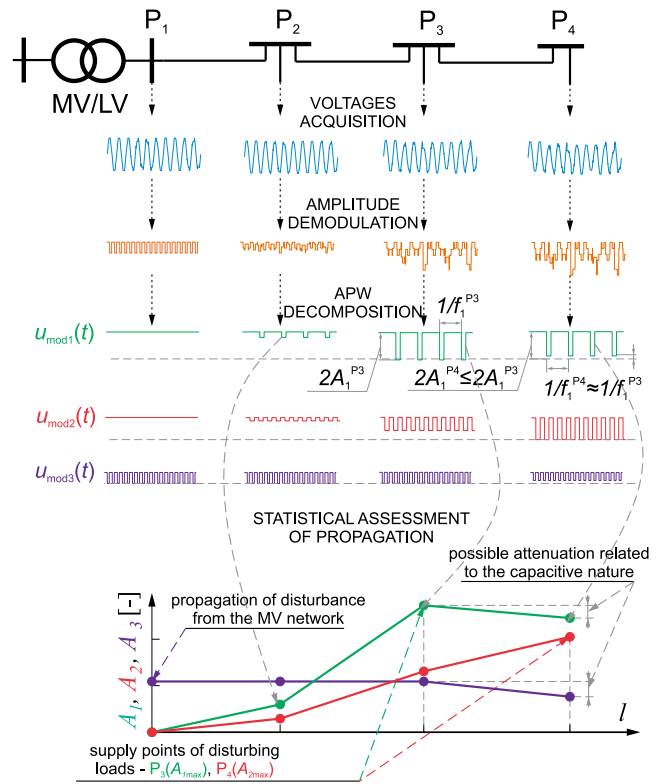

Fig. 3. The conceptual operation of the proposed approach.

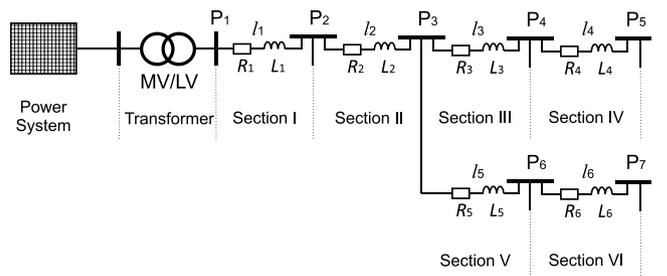

Fig. 4. The LV power network configurations selected in the research.

To sum up, in order to automatically identify sources of voltage fluctuations without additional expert knowledge, the procedure presented in Fig. 1 should be implemented in the infrastructure of smart energy meters in the power grid. Individual meters in subsequent blocks of a specific duration, e.g. 10 min (the duration of the block can be any, the proposal of 10 min implies that it is the time for determining the indicator $P_{st}$ currently used to assess the flicker severity), acquire the voltage $u(t)$. If the limit value for the selected control parameter (e.g., value equal to 1.0 for the indicator $P_{st}$) has been exceeded on any meter in a given network with a common topology, the procedure for identifying the sources of voltage fluctuations is carried out in individual meters. For this purpose, a process defined by the signal chain is performed on individual meters: demodulation [46] – decomposition [47] – statistical assessment of the propagation of voltage fluctuations.

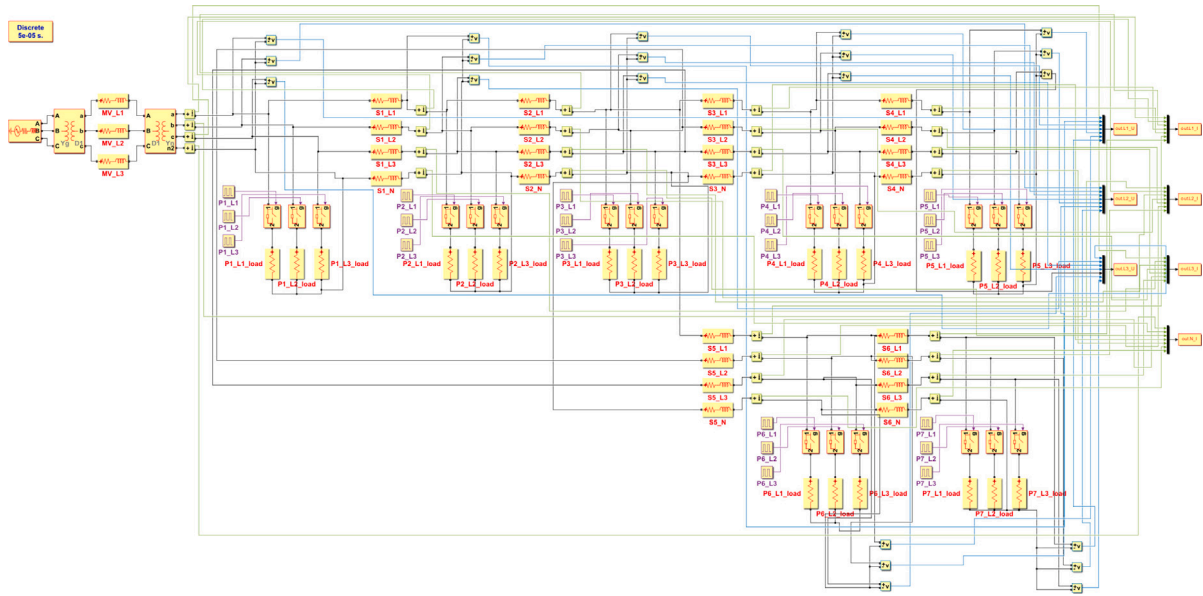

**Fig. 5.** The model of the configuration of the power grid implemented in MATLAB/SIMULINK for the example case.

## 3. Research results and discussion

### 3.1. Numerical simulation studies

Numerical simulation studies were carried out using a three-phase power network model with a branch radial topology, the diagram of which is shown in Fig. 4. The power network model was implemented in the SIMULINK/MATLAB program as shown in Fig. 5 for the exemplary case analyzed. The research focused mainly on LV networks, because the voltage fluctuations most often occur in them. Currently, most of the LV networks in Poland are radial topologies with branches. As part of the numerical simulation studies, a test set was prepared, consisting of 1000 cases in which the following were randomly adopted:

- different values of the parameters of the adopted power network model, where $R_i \in [50; 150]$ m$\Omega$, $L_i \in [6.7; 100]$ µH;
- a different number $N$ of disturbing loads (both single-phase and three-phase), where $N \in [2; 6]$, with different parameters such as the frequency of changes in the operating state of disturbing loads $f_i \in (0.1; 150)$ Hz and the active power of individual sources of disturbance $P_{N_i} \in (0.5; 4)$ kW;
- different supply points $P_i$ of individual disturbance sources.

Detailed values of individual parameters for the particular cases considered are included in a publicly available database [48]. Considering the randomly adopted values of particular parameters in numerical simulation studies allows an effective assessment of the usefulness of the proposed solution. As part of numerical simulation studies, for individual cases voltages were acquired from particular points of the power network with a sampling rate of 20 kSa/s, and then an identification procedure was carried out for the data obtained in this way, focused on two objectives:

- the indication of the supply point of disturbing loads;
- the estimation of the frequency of changes in the operating state of individual sources of disturbance (in practice, this process significantly supports the localization of sources of disturbance, even using methods other than the proposed one — e.g. single-point methods).

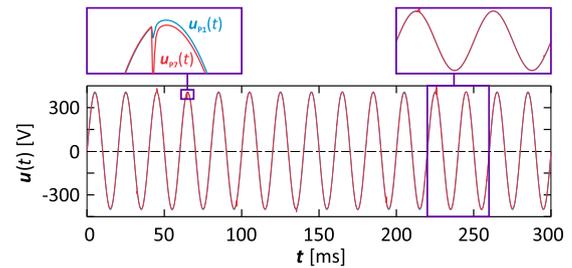

**Fig. 6.** The acquired voltage signal for the selected case in the L1 phase for network points P1 and P7 — numerical simulation studies.

**Table 1**
Parameters of the real power network model determined at a power frequency of 50 Hz.

| Section | $R$ [m$\Omega$] | $X$ [m$\Omega$] | $Z$ [m$\Omega$] |
|---|---|---|---|
| MV/LV transformer | 3.8 | 10.8 | 11.4 |
| Section I | 150.0 | 31.4 | 153.3 |
| Section II | 150.0 | 31.4 | 153.3 |
| Section III | 100.0 | 69.1 | 121.6 |
| Section IV | 50.0 | 2.1 | 50.0 |
| Section V | 100.0 | 69.1 | 121.6 |
| Section VI | 50.0 | 2.1 | 50.0 |

Fig. 6 shows the acquired voltage signal for the selected case of numerical simulation studies in the L1 phase for network points P1 and P7.

### 3.2. Laboratory experimental studies

Laboratory experimental studies were carried out using a three-phase real power grid three-phase model specially prepared for research purposes, which is shown in Fig. 7. The block diagram of the laboratory setup is shown in Fig. 8. The real power grid model is supplied directly from a 630 kVA MV/LV transformer. The diagram of prepared real power grid model is analogous to the diagram presented in Fig. 4, for which the appropriate values presented in Table 1 are adopted.

As part of the laboratory experimental studies, 100 cases were considered, in which the following randomly adopted parameters were considered:

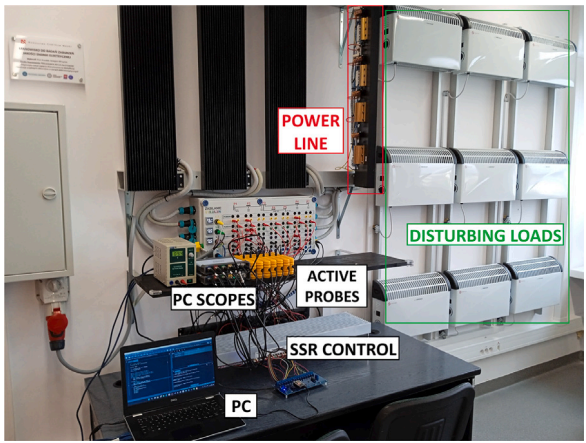

**Fig. 7.** The view of the specially prepared real model of the power network for research purposes.

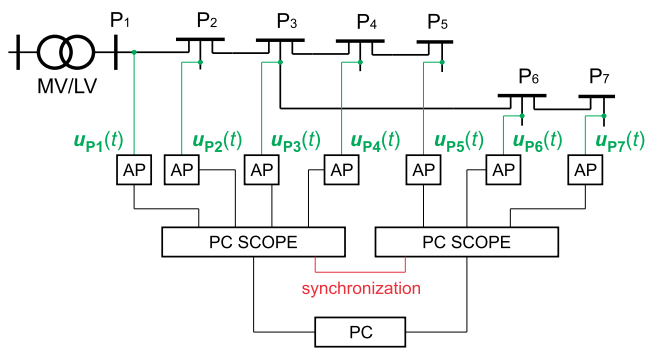

**Fig. 8.** The diagram of the specially prepared real model of the power network, where AP is the active probe Pico TA041 and PC SCOPE is the virtual oscilloscope PicoScope 5444D.

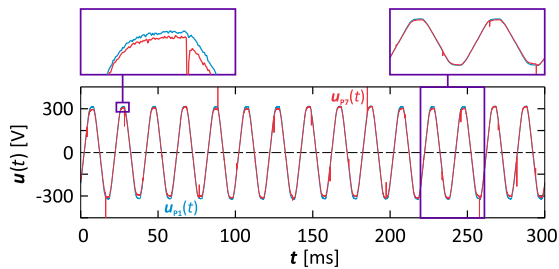

**Fig. 9.** The acquired voltage signal for the selected case in the L1 phase for network points P1 and P7 — laboratory experimental studies.

- a different number $N$ of disturbing loads (both single-phase and three-phase), where $N \in \langle 2; 6 \rangle$, with different the frequencies of changes in the operating state of disturbing loads $f_i \in \langle 0.1; 150 \rangle$ Hz with the active power of individual sources of disturbance $P_{N_i} = 2$ kW;
- different supply points $P_i$ of individual disturbance sources.

Fig. 9 shows the acquired voltage signal for the selected case of laboratory experimental studies in the L1 phase for network points P1 and P7.

Disturbing loads causing voltage fluctuations in the research were 2 kW convection systems controlled by solid state relay (SSR) based on MOSFET switches. Detailed values of individual parameters for the particular cases considered are included in a publicly available database [48]. As part of laboratory experimental studies, voltages were acquired for individual cases from particular points of the power network with a sampling rate of 20 kSa/s. The database [48] contains 100 sets of recorded voltages from individual points, which were used in the investigation. Then, for the data obtained in this way, the identification procedure was carried out focused on two objectives:

- the indication of the supply point of disturbing loads;
- the estimation of the frequency of changes in the operating state of individual sources of disturbance (in practice, this process significantly supports the localization of sources of disturbance, even using methods other than the proposed one — e.g. single-point methods).

### 3.3. Research results

During the research, for data from numerical simulation studies and laboratory experimental studies, an identification procedure was performed using the proposed approach (PA) and selected other methods available in the literature:

- method based on the assessment of the propagation of voltage fluctuations using non-parametric statistical assessment of voltage fluctuation indices without additional expert knowledge (M1) [30,31];
- method based on the evaluation of flicker propagation (M2) [33, 49];
- method based on decomposition using enhanced empirical wavelet transform (M3) [38].

It is worth noting that the data used to verify the proposed method recreate the actual conditions that occur in the modern power grid (e.g. simultaneous occurrence of voltage fluctuations and distortions).

Two parameters are used to evaluate the performance of the identification of disturbing loads. The first parameter $e_1$ is the accuracy metric that determines the effectiveness of a correct indication of the supply point of the source of disturbance. The parameter $e_1$ is given by:

$$e_1 = \frac{N_T}{N} \cdot 100\%, \quad (1)$$

where $N_T$ is the number of correctly identified supply points of disturbance sources, and $N$ is the actual number of considered disturbance sources in individual cases. The second parameter $e_2$ is the accuracy metric that determines the effectiveness of the correct estimation of the frequency of changes in the operating state of individual sources of voltage fluctuations. The parameter $e_2$ is given by:

$$e_2 = \frac{N_{f_T}}{N} \cdot 100\%, \quad (2)$$

where $N_{f_T}$ is the number of correctly estimated frequencies of changes in the operating state of individual sources of disturbance with an accuracy of 10%, and $N$ is the actual number of considered sources of disturbance in individual cases.

Table 2 shows the determined parameters $e_1$ and $e_2$ for the proposed approach and other selected methods of identifying disturbing loads for data from numerical simulation studies and laboratory experimental studies. When analyzing the results of the research obtained, the usefulness of the proposed approach can be clearly determined. Fig. 10 shows an example of the operation of the proposed approach on data obtained from experimental studies for the selected phase L1. In this selected case, three disturbing loads from different points on the line were supplied from the phase L1 of the tested power line.

Analyzing the obtained research results, it can be noticed that, apart from the proposed method PA, there is a visible tendency that the identification process based on data obtained in the real power grid has lower effectiveness (more frequent incorrect indication of the supply point of disturbing loads and more frequent incorrect estimation of the frequency of changes in the operating state of the disturbing load) than identification process based on data from numerical simulation tests. This tendency is the result of the fact that in the real power grid, in addition to the disturbance caused by known specific sources of disturbance used in the laboratory, there were other disturbances (e.g. distortion of the supply voltage by higher harmonics or small voltage fluctuations related to the influence of other loads) propagated from the MV side of power network or from other circuits supplied from the same MV/LV switchboard. The proposed solution PA allows for effective localization of disturbing loads in the power grid (correct indication of the supply points of the disturbing loads) and the proposed solution PA provides the most accurate estimate of the frequency of changes in the operating state of disturbing loads, which parameter significantly supports the process of localization of disturbance sources in the power grid. The effectiveness of identification of sources of disturbances by the proposed method PA is maintained both for numerical simulation studies and for experimental laboratory studies, and therefore the proposed approach PA is not sensitive to other occurring simultaneously low-frequency disturbances in the power grid. Minor errors in estimation of the frequency of changes in the operating state of disturbance sources using the proposed approach PA in numerical simulation studies are related to errors in the demodulation method used [46] (as a result of the statistical filter used in this demodulator, as described in [46]), due to which the demodulator can distort the estimated modulating signal in the frequency range close to the cut-off frequency of the statistical filter used in the demodulator. However, it is acceptable because first, the use of the indicated demodulator allows considering loads that change their operating state with a frequency higher than the power frequency, and second, despite the errors caused by the demodulator used, the use of the proposed decomposition method [47] allows an accurate indication of the supply points for individual sources of disturbance. The disturbance that occurs is proportional to individual points in the power grid, which in turn does not disturb the process of evaluation of the propagation of the signal components used in the process of localization (indication of supply points) of sources of disturbance. The lack of higher errors in estimation of the frequency of changes in the operating state of disturbance sources by the proposed method PA in laboratory experimental studies is the result of the smaller size of the verification set (smaller number of cases considered, which can disturb the identification process).

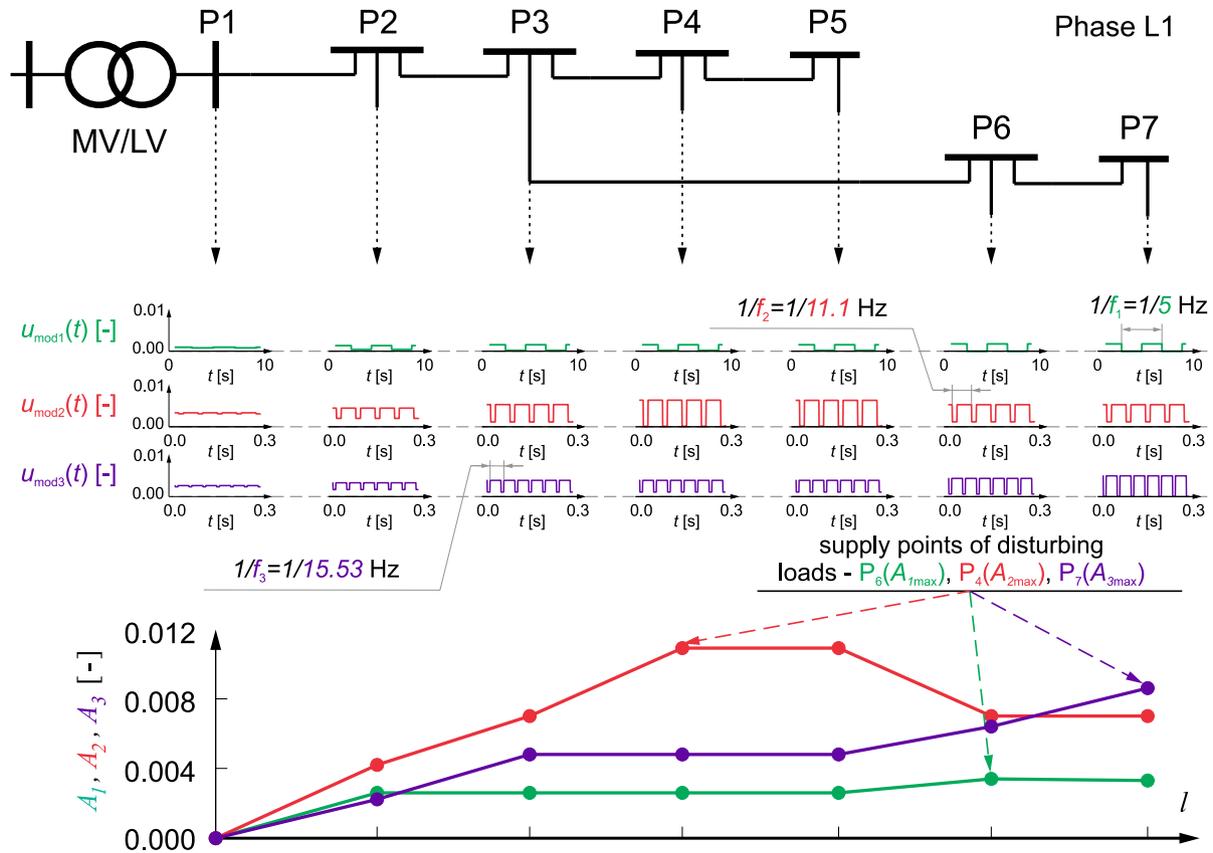

**Fig. 10.** The example of the operation of the proposed approach on data obtained from experimental studies for the selected phase L1.

**Table 2**
Determined parameters $e_1$ and $e_2$ for the proposed approach and other selected methods for data from numerical simulation studies (simulations) and laboratory experimental studies (experiments).

| Method | $e_1$ [%] | | $e_2$ [%] | |
|---|---|---|---|---|
| | Simulations | Experiments | Simulations | Experiments |
| PA | 100.00 | 100.00 | 92.52 | 100.00 |
| M1 | 67.59 | 43.49 | 91.01 | 88.09 |
| M2 | 85.98 | 69.53 | 91.68 | 93.35 |
| M3 | 53.50 | 46.81 | 81.04 | 82.27 |

## 4. Conclusion

The article presents a proposed algorithm for the identification of sources of voltage fluctuations in the radial power network, which allows for the determination of selected parameters associated with the impact of individual disturbing loads, based on which it is possible to indicate the point of supply of individual disturbing loads. The selected parameters are the frequency of changes in the operating state of individual sources of disturbance and the amplitude of the voltage fluctuations caused by them, associated with the specific contribution of the source of disturbance in the resulting voltage fluctuations. The presented approach considers sources of disturbance that change their operating state at a frequency both lower and higher than the power frequency. Particularly innovative is the inclusion of disturbing loads that change their operating state at a frequency higher than the power frequency. Such disturbing loads are, for example, power electronic devices, the number of which in the power grid is increasing. The presented algorithm is an improvement of the procedure based on the enhanced empirical wavelet transform, which allows for the elimination of limitations resulting from the decomposition method used. The correctness of the proposed approach was verified in experimental and simulation studies. The research considers the branched radial topology common in LV networks. The research includes sections with different impedance and line character that can cause incorrect identification (including localization) of disturbance sources using currently used statistical methods. This problem does not affect the accuracy of the proposed approach, and therefore the identification process is performed most effectively by the proposed algorithm. The only minor errors associated with the proposed solution are related to the demodulation method used, in which bandwidth limitation can affect the accuracy of estimation of selected parameters associated with individual sources of disturbance due to bandwidth limitation by the statistical filter used in the indicated demodulator.

**CRediT authorship contribution statement**

**Piotr Kuwałek:** Writing – review & editing, Writing – original draft, Visualization, Validation, Software, Resources, Project administration, Methodology, Investigation, Funding acquisition, Formal analysis, Data curation, Conceptualization. **Grzegorz Wiczyński:** Supervision.

**Declaration of competing interest**

The authors declare the following financial interests/personal relationships which may be considered as potential competing interests: Piotr Kuwalek reports financial support was provided by National Science Centre. If there are other authors, they declare that they have no known competing financial interests or personal relationships that could have appeared to influence the work reported in this paper.

**Data availability**

Data will be made available on request.

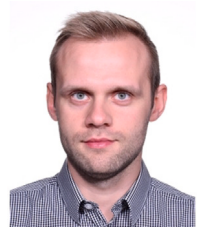

**Piotr Kuwałek** received the B.Sc. degree in mathematics, M.Sc. Eng. in electrical engineering, and Ph.D. degrees in automation, electronic and electrical engineering from the Poznan University of Technology, Poznan, Poland, in 2018, 2018 and 2022, respectively. He is currently an Assistant Professor with the Division of Metrology, Electronics and Lighting Engineering at Poznan University of Technology. His current research interests include power quality evaluation and signal processing.

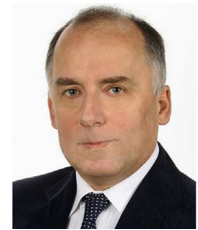

**Grzegorz Wiczyńki** received the M.Sc. and Ph.D. degrees in electrical engineering from the Poznan University of Technology, Poznan, Poland, in 1990 and 1998, respectively. He is currently with the Division of Metrology, Electronics and Lighting Engineering, Poznan University of Technology. His current research interests include electronics and optoelectronics in electrical and nonelectrical quantities measurements and power quality evaluation.